\documentstyle[12pt,aaspp4]{article}
\begin{document}

\title{MACHO Mass Determination Based on Space Telescope Observation}
\author{Mareki Honma\altaffilmark{1}}
\affil{Institute of Astronomy, University of Tokyo, Mitaka, 181-8588, Japan}
\authoremail{honmamr@cc.nao.ac.jp}
\altaffiltext{1}{Research Fellow of the Japan Society for the Promotion of Science}
\slugcomment{submitted to the Astrophysical Journal Letter}

\begin{abstract}
We investigate the possibility of lens mass determination for a caustic crossing microlensing event based on a space telescope observation.
We demonstrate that the parallax due to the orbital motion of a space telescope causes a periodic fluctuation of the light curve, from which the lens distance can be derived.
Since the proper motion of the lens relative to the source is also measurable for a caustic crossing event, one can find a full solution for microlensing properties of the event, including the lens mass.
To determine the lens mass with sufficient accuracy, the light curve near the caustic crossing should be observed within uncertainty of $\sim$ 1\%.
We argue that the Hubble Space Telescope observation of the caustic crossing supplied with ground-based observations of the full light curve will enable us to determine the mass of MACHOs, which is crucial for understanding the nature of MACHOs.
\end{abstract}


\section{Introduction}

Recent extensive searches for gravitational microlensing events have already detected about a hundred of events toward the Galactic bulge and the Magellanic Clouds (Alcock et al. 1993; 1996; 1997a; 1997b; Aubourg et al.1993; Alard \& Guibert 1997; Udalski et al. 1994).
The number of events toward the Magellanic clouds exceeds the expected number of events for known population of stars in the Galactic disk and the Magellanic Clouds themselves, indicating that a considerable number of the events are caused by massive compact halo objects (hereafter MACHOs).
However, since the MACHO mass cannot be obtained directly from microlensing observations, the nature of MACHOs remains unclear.
The only observable for a single microlensing event is the Einstein ring crossing time $t_{\rm E}$, which may be written as,
\begin{equation}
\label{eq:E-crossingtime}
t_{\rm E} = \frac{1}{v_\perp}\sqrt{\frac{4Gm}{c^2} D_{\rm s} l (1-l)}
\end{equation}
where $v_\perp$ is the tangential velocity of the lens, $m$ is the lens mass, $D_{\rm s}$ is the source distance, and $l$ is the fractional lens distance, i.e., the ratio of the lens distance to the source distance.
Since the right-hand side of equation (\ref{eq:E-crossingtime}) contains three unknowns ($v_\perp$, $l$ and $m$), the MACHO mass cannot be determined for each microlensing event.
Instead, the MACHO mass is evaluated statistically by assuming a halo model which describes the distribution of the lens distance and velocity.
Alcock et al.(1997b) found the MACHO mass of $\sim$ 0.5 $M_\odot$ by assuming the standard halo model in which the rotation curve is flat out to the Magellanic Clouds, and suggested that MACHOs are likely to be old white dwarfs.
However, the MACHO mass may be significantly changed when a different halo model is considered.
In fact, if the rotation curve is slightly declining in the outer region, the MACHO mass is as small as that of brown dwarfs (Honma \& Kan-ya 1998).

While microlensing events by single lenses cannot break the three-fold degeneracy in equation (\ref{eq:E-crossingtime}), one can additionally measure the proper motion for a caustic crossing event, which is a microlensing event due to a binary lens system (e.g., Schneider \& Weiss 1986).
To measure the proper motion, a real-time detection of the event is necessary because the caustic crossing must be monitored intensively with high time resolution.
Recent developments of the alert system made the real-time detection possible, and in fact, the proper motion has been measure for the first time for the event 98-SMC-01 (Afonso et al. 1998; Albrow et al. 1998; Alcock et al. 1998; Rhie et al. 1998).
Once the proper motion is measured, the only quantity necessary for the lens mass determination is the lens distance.
Hardy \& Walker (1995) showed that if a caustic crossing event is monitored at two or more observatories that are separated well, the parallax effect causes a time delay of $\sim$ 30 sec, from which the lens distance can be derived.
Later, Gould \& Andronov (1998) discussed extensively the possibility of lens mass determination by combination of the proper motion and parallax measurements for a caustic crossing event.
For doing this, Gould \& Andronov (1998) proposed a monitoring observation of a caustic crossing event with three non-collinear telescopes that are located in different continents.
This proposal requires that the event should be observable from three continents, i.e., three continents must be in the dark side of the Earth at the same time, and the weather must be clear.
This requirement may strongly limit the application of this method.

As an alternative to three ground-based telescopes, in this Letter we propose to use a space telescope for lens distance determination.
Because its orbital motion automatically causes parallax effect, a space telescope acts like a number of ground-based telescopes located in difference continents.
Also, there is no concern for the weather in the space.
Therefore, using a space telescope is more practical and realistic than using three ground-based telescopes.
In the following sections, we describe how parallax effect changes the light curve of a caustic crossing event, and discuss how accurately one can determination the lens distance as well as the lens mass based on a space telescope observation.

\section{Parallax from Space Telescope}

Here we describe the parallax effect in a caustic crossing event observed with a space telescope.
In the following analysis, the origin of the coordinate system is set to be the center of the Earth, and $z$ axis is set to be in the direction of the source star.
The $x$ axis is set to be perpendicular to both of the $z$ axis and the orbital axis of the space telescope (thus the orbital axis is in the $y$-$z$ plane).
We define the inclination of the telescope orbit as an angle between the $z$ axis and the orbital axis.
We also assume that the space telescope is in a circular orbit with a radius of $r_{\rm st}$ and an angular velocity of $\omega$.
The position of the telescope ($\vec{T}$) is written as
\begin{equation}
\label{eq:telescope-position}
\vec{T} = (r_{\rm st} \cos (\omega t+\delta), r_{\rm st} \sin (\omega t+\delta) \cos i, r_{\rm st} \sin (\omega t+\delta) \sin i).
\end{equation}
We set $t=0$ when the source observed from the center of the Earth crosses the caustic.
The angle $\delta$ describes the position of the telescope at $t=0$.

In this coordinate system the Earth and the source are at rest and the caustic is moving.
The position of the caustic crossing point (the point on the caustic where the source crosses when observed from the center of the Earth) is given by,
\begin{equation}
\vec{C_{\rm l}}=(v_\perp t \cos \alpha, v_\perp t \sin \alpha, D_{\rm d}+v_{\rm r}t),
\end{equation}
where $v_\perp$ and $v_{\rm r}$ are tangential and radial velocity of the lens, and $\alpha$ is the angle between the $x$ axis and the direction of tangential velocity $v_\perp$.
When observed from the center of the Earth, the position of the caustic crossing point projected onto the source plane is given by,
\begin{equation}
\vec{C_{\rm s}}=\frac{D_{\rm s}}{D_{\rm d}}\vec{C_{\rm l}} \equiv \frac{1}{l}\vec{C_{\rm l}}.
\end{equation}
When observed from a space telescope orbiting the Earth, the projected position of the caustic crossing point in the source plane is given by,
\begin{equation}
\vec{C_{\rm s}'}=\vec{T}+\frac{1}{l}(\vec{C_{\rm l}}-\vec{T}).
\end{equation}

For convenience we define the source position relative to the caustic crossing point in the source plane as $\vec{X}= \vec{S}-\vec{C_{\rm s}}$ and $\vec{X'}= \vec{S}-\vec{C_{\rm s}'}$, where $\vec{S}$ is the position of the source, and $\vec{S}=(0, 0, D_{\rm s})$.
With these equations, we may write $\vec{X}$ and $\vec{X'}$ explicitly as
\begin{equation}
\label{eq:X-explicit}
\vec{X}=\left(-\frac{v_\perp t}{l}\cos\alpha, -\frac{v_\perp t}{l}\sin\alpha, 0\right),
\end{equation}
\begin{equation}
\label{eq:X'-explicit}
\vec{X'}=\left(-\frac{v_\perp t}{l}\cos\alpha + \frac{1-l}{l}r_{\rm st}\cos(\omega t+\delta), -\frac{v_\perp t}{l}\sin\alpha + \frac{1-l}{l}r_{\rm st}\sin(\omega t+\delta)\cos i, 0\right)
\end{equation}
Equation (\ref{eq:X'-explicit}) shows that the parallax effect due to the telescope motion causes a wavy trajectory of the source relative to the caustic (note that in the coordinate system of $X_x$ and $X_y$ the source is moving while the caustic is at rest).

\section{Lens Distance from Light Curve Observation}

Here we calculate the light curve near the caustic observed with a space telescope.
For simplicity we assume a constant surface brightness of the source star.
Gould \& Andronov (1998) obtained the magnification for such a source near the caustic as, 
\begin{equation}
A=a_0 G(\eta) + a_1,
\end{equation}
\begin{equation}
G(\eta)=\frac{2}{\pi}\int^{1}_{\max(\eta,1)} \left(\frac{1-x^2}{x-\eta}\right)^{1/2} dx,
\end{equation}
where the function $G$ describes the shape of light curve, and $\eta$ is the separation between the source and the caustic normalized with the source size $R_*$, namely, $\eta\equiv d/R_*$.
The two constant $a_0$ and $a_1$ describe the maximum magnification and the magnification outside the caustic, respectively.
Note that these two constants depend on the shape of the caustic, the source trajectory, the source radius, and so on.

To evaluate $\eta$ as a function of time, here we assume that the caustic is approximately linear near the caustic crossing point.
This approximation is valid for stellar or substellar mass lenses in the halo because the amplitude of wavy motion of the source due to the parallax is usually much smaller than the size of caustic itself.
In the coordinate system of $X_x$ and $X_y$ (see the definition of $\vec{X}$), the caustic is expressed as $X_y = \tan (\alpha + \phi) X_x$, where $\phi$ is the angle between the caustic and the direction of the source motion.
In the absence of the parallax effect, the distance of the source to the caustic is given by
\begin{equation}
\eta = \left(\sin(\alpha + \phi) X_x - \cos(\alpha+\phi) X_y \right)/R_*.
\end{equation}
Similarly, when observed from a space telescope orbiting around the Earth, the distance of the source from the caustic is obtained as 
\begin{equation}
\label{eq:eta}
\eta' = \left(\sin(\alpha + \phi) X_x' - \cos(\alpha+\phi) X_y' \right)/R_*.
\end{equation}
Note that $\eta'$ depends on following parameters : $\alpha$, $\phi$, $v_\perp$, $l$, $R_*$, $r_{\rm st}$, $\omega$, $\delta$, $i$, and $t$.
Among them, the unknown parameters to be determined by space telescope are $l$ and $\alpha$, i.e., the lens distance and the direction of the source motion.
The orbital parameters of the telescope ($r_{\rm st}$, $\omega$, $\delta$, $i$) are accurately known.
One can also determine the angle $\phi$ and the proper motion $\mu$ from the full light curve fitting of the caustic crossing event.
To obtain the full light curve, the ground-based observation is necessary because of long event duration ($t_{\rm E}\sim 40$ days).
If the lens is located in the halo and the source proper motion can be neglected, the proper motion $\mu$ is simply related to $v_\perp$ as $v_\perp = l D_{\rm s} \mu$, and hence $v_\perp$ can be determined once $l$ is given.
The radius of the source star $R_*$ can be obtained based on the color and the spectrum of the source star.
The accuracy of $R_*$  is typically within 10\%, but depends on how precisely one can determine the effective temperature.
An accurate measurement of $R_*$ is crucial for the lens mass determination because $R_*$ affects not only the lens distance $l$ through equation (\ref{eq:eta}) but also the proper motion $\mu$ derived from the full light curve.

Figure 1 shows $G(\eta(t))$ and $G(\eta'(t))$ as well as the difference of the two, $\delta G\equiv G(\eta(t))-G(\eta'(t))$.
In figure 1 we assume $D_{\rm s}=50$ kpc, $l=0.2$, $v_\perp=100$ km/s, $\alpha=20^\circ$, $\phi=45^\circ$ and $R_*=3R_\odot$, corresponding to a typical microlensing event due to MACHOs in the Galactic halo (e.g., Paczynski 1986).
With these parameters, it takes about 3 hours for the source to cross the caustic.
For orbital parameters for the space telescope, we assume $r_{\rm st}=7000$ km, $P_{\rm orb}\equiv 2\pi/\omega = 97$ minutes, $\delta = 160^\circ$ and $i=30^\circ$, which are similar to those of the Hubble Space Telescope.
Figure 1 shows the light curve near the caustic for about 5 hours, corresponding to $\eta=-2$ to $1$.
The figure demonstrates that the parallax effect due to the telescope motion causes the periodic fluctuation in $\delta G$.
Note that changing the angle $\alpha$ mainly shifts the phase of the $\delta G$ curve whereas the lens distance $l$ changes the amplitude of the curve.
Thus, both $l$ and $\alpha$ can be determined by fitting the $\delta G$ curve.
Since the amplitude of the $\delta G$ curve is of a few \%, the source magnification should be measured within uncertainty of 1 \%, or with a photometric S/N larger than $\sim 100$.
For a space telescope like the HST, this S/N is easily achievable when the source is being magnified significantly.

\section{Uncertainty in Lens Distance and Mass}

In this section, we investigate how accurately we can measure the parameters $l$ and $\alpha$ as well as the lens mass $m$.
First, we evaluate how the lens distance uncertainty depends on the photometric S/N.
For doing this, we have simulated observations of the light curve presented in figure 1.
We assumed that the light curve was observed every 10 minutes with a constant uncertainty.
The lens and orbital parameters are set to be the same to those in section 3 (simulated observations with the photometric $S/N$ of 300 are plotted in figure 1).
We calculated the likelihood in the parameter space of $l$ and $\alpha$ as $L = \prod p(\delta G_i | l, \alpha)$, where $\delta G_i$ corresponds to the $i$-th simulated measurement of the light curve, and the probability $p$ is calculated assuming Gaussian error.
We then calculated best values of $l$ and $\alpha$ as well as their uncertainties $\Delta l$ and $\Delta \alpha$ (68\% confidence level).
Figure 2 plots the resultant uncertainties in $l$ and $\alpha$ with S/N ratio of 50 to 300.
Figure 2 shows that the lens distance is determined within 30\% if the photometric S/N ratio is greater than 100.
In this case, one can also determine the direction of the lens motion within 20 degrees.
Figure 2 also shows that the uncertainties decrease gradually with increasing the photometric S/N ratio, and that the lens distance can be determined within $\sim 10$\% in case of S/N $\sim$ 300.

Next, in order to investigate how far we can measure the lens distance, we have also calculated the uncertainties in $l$ and $\alpha$ with varying $l$ while keeping the S/N constant.
We assumed that the lens path relative to the caustic is the same to that in figure 1, and also assumed that the proper motion $\mu$ is $10.0$ km/s/kpc, so that all the events considered here have the same light curve to the one presented in figure 1 regardless of the lens distance $l$.
Figure 3 plots the uncertainties in $l$ and $\alpha$ with varying $l$ for every 10 minutes observations with the photometric S/N of 300.
Figure 3 demonstrates that the uncertainty in $l$ is rapidly increasing with the lens distance.
Nevertheless, the distance uncertainty remains less than 30 \% for lenses within $l=0.5$.
Since typical microlensing events by MACHOs have $l\sim 0.2$ and since most of microlensing events by MACHOs have $l$ less than 0.5 (Paczynski 1986), the lens distance can be determined for most of events.

Note that the uncertainty in $R_*$, which was not considered above, may not be negligible in some cases.
Since the uncertainty in $R_*$, which is typically within 10\%, propagates to the uncertainty in $l$ almost linearly, its effect may be important for an event with $\Delta l/l$ less than $\sim 10$\%.
Also, a possible discontinuity of the space telescope observation due to the occultation by the Earth could affect the lens distance uncertainty.
If the peaks of $\delta G$ curve are missed due to the occultation, the uncertainty in $l$ may increase considerably.
However, for instance, the Magellanic Clouds is in the Continuous Viewing Zone of the Hubble Space Telescope.
Thus, a space telescope which has an orbit close to the HST can perform a continuous or nearly continuous observation depending on the orbital precession, and so the probability of missing the peaks of $\delta G$ curve will be small.

Once $l$ is determined, the MACHO mass can be obtained through equation (\ref{eq:E-crossingtime}).
For a binary lens, the mass $m$ in equation (\ref{eq:E-crossingtime}) denotes the total lens mass, but the mass ratio of the two lenses can be also derived from the full light curve.
Since the proper motion $\mu$ and the Einstein ring crossing rime $t_{\rm E}$ can be measured relatively accurately from the full light curve, the accuracy of the MACHO mass depends mainly on the accuracy of $l$.
Thus, if S/N of a few hundred is achieved, the MACHO mass can be determined with an uncertainty of 10 $\sim$ 30 \%.
If $\Delta l$ is larger, the constraint on the MACHO mass may be weak.
However, even in that case, a measurement of the lens distance allow us to discriminate whether the lens is in the halo or not, which will be a strong test for existence of MACHOs.

\section{Discussion}

We have seen that a space telescope observation of a caustic crossing event supplied with the ground-based observations will enable us to measure the MACHO mass.
The most practical strategy at present is: 1) real-time detection and long-term monitoring of a caustic crossing event by ground-based observations, and 2) the space telescope observation of the caustic crossing.
The real-time detection of the event is essential, and this can be done by the existing alert system.
For the lens trajectory determination from the full light curve, precise photometry with high time resolution as well as global collaboration of monitoring groups will be efficient.
High time resolution is also crucial for predicting the precise date of the second caustic crossing, which is necessarily for scheduling the observation with a space telescope.
Since a typical interval of two caustic crossings is less than 10 days for binary MACHOs (Honma 1999), monitoring the source twice or more per night is more favorable than the nightly observation.

If the fraction of binary MACHOs is as high as that of binary stars, $5\sim 10$ \% of events are expected to be a caustic crossing event (Mao \& Paczynski 1991).
Thus, one can expect a caustic crossing event per a few year even with the current microlensing search, and if the next generation microlensing search is launched, the expected number of such an event will be increased by two order of magnitudes (Stubbs 1998).
Therefore in a decade or so, we may be able to measure the MACHO mass for a number of events by combining the ground-based monitoring and the HST or other space telescope in the next generation.

\acknowledgments

The author acknowledges the financial support from the Japan Society of the Promotion of science.

\clearpage

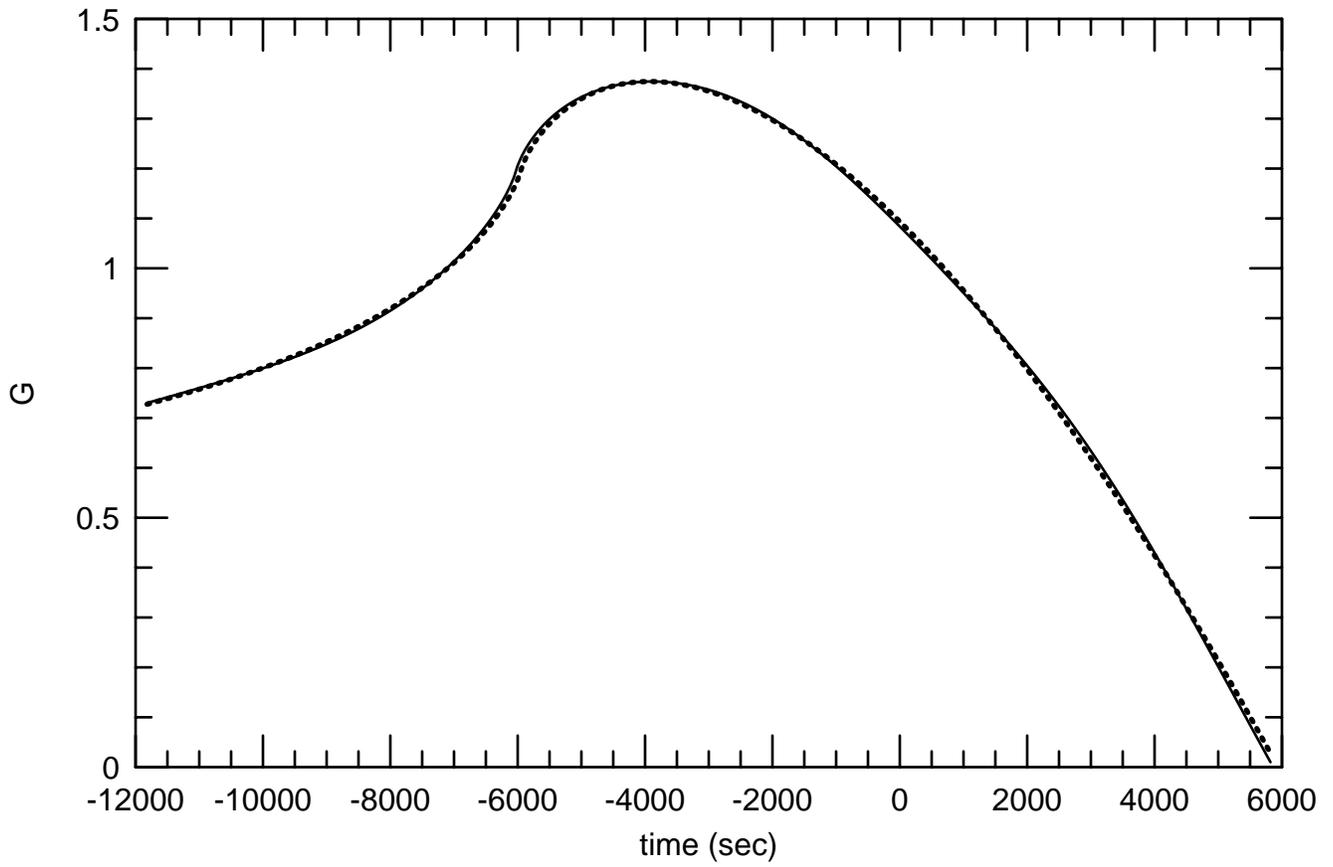
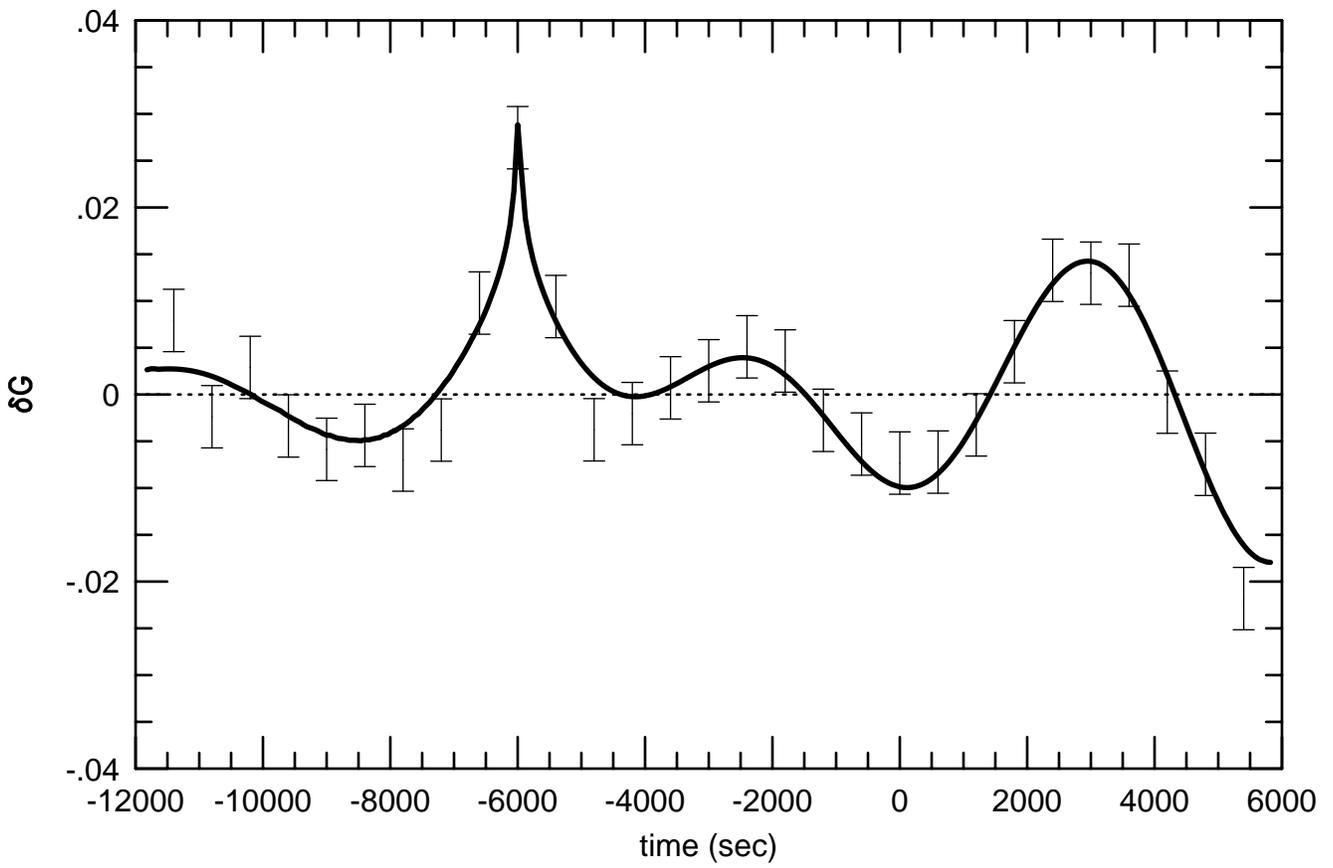
\begin{figure}
\caption{Upper panel shows the time variation of $G$ seen from the Earth (dotted line) and the space telescope (solid line).
Lower panel is the time variation of $\delta G$ (the difference of the two $G$ curves).
Error bars are simulated observations with the interval of 10 minutes and the photometric S/N of 300.}
\end{figure}

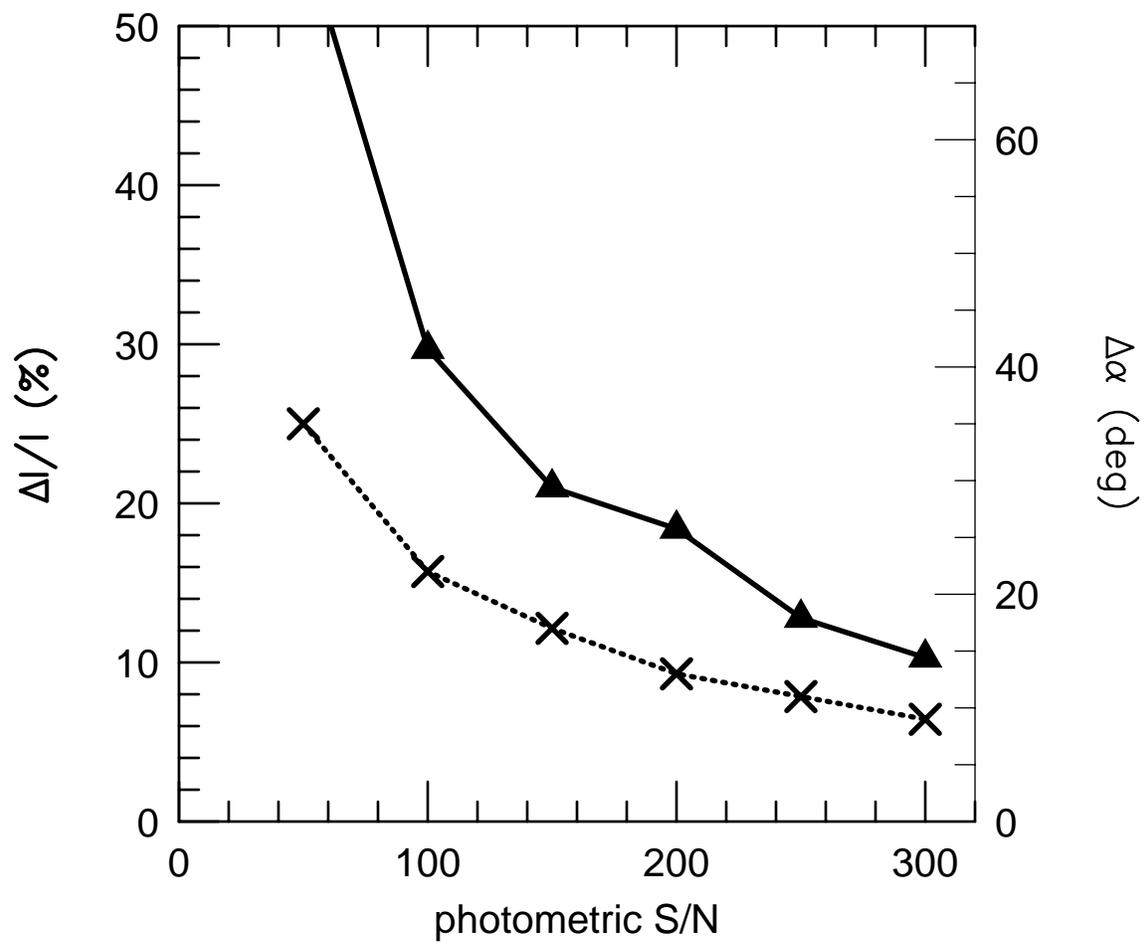
\begin{figure}
\caption{Uncertainty in $l$ and $\alpha$ with the photometric S/N.
The lens distance is assumed to be $l=0.2$.
The triangles are for the lens distance uncertainty ($\Delta l/l$), and crosses are for the uncertainty in the direction of the source motion ($\Delta \alpha$).}
\end{figure}

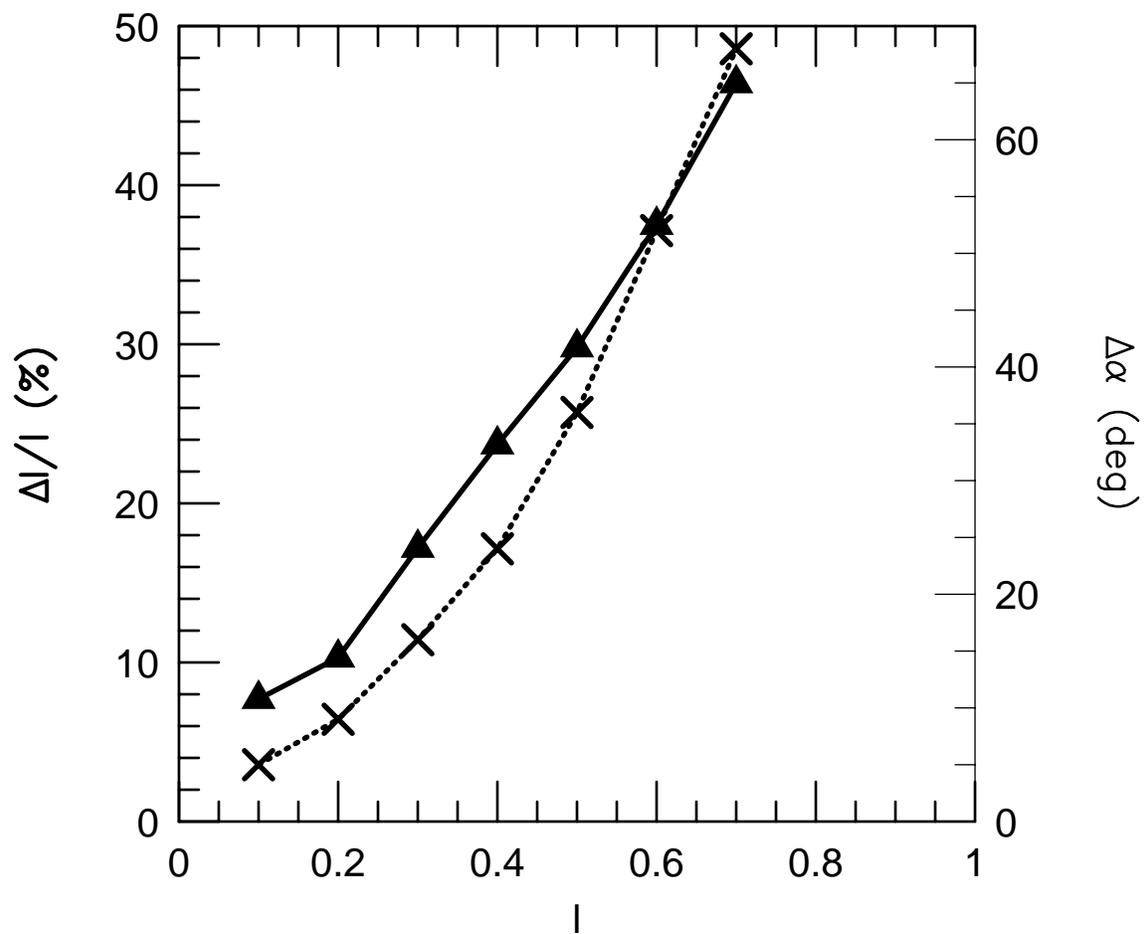
\begin{figure}
\caption{Uncertainty in $l$ and $\alpha$ with the lens distance $l$.
The photometric S/N is assumed to be 300.
The symbols are the same to those in figure 2.}
\end{figure}

\end{document}